\documentclass[11pt,a4paper]{amsart}
\usepackage[dvips,colorlinks=true,linkcolor=blue,urlcolor=red]{hyperref}
\usepackage[english]{babel}
\usepackage{graphicx,subfigure} 
\usepackage{wrapfig}
\usepackage{stmaryrd}
\usepackage{amssymb,amsfonts}
\numberwithin{equation}{section}

\newcommand{\car}{\mathbf{1}}
\newcommand{\R}{\mathbb{R}}

\newcommand{\N}{\mathbb{N}}

\newcommand{\Z}{\mathbb{Z}}

\newcommand{\C}{\mathbb{C}}

\newcommand{\vers}{\operatornamewithlimits{\to}}

\newcommand{\D}{\displaystyle}

\newcommand{\pro}{\mathbb{P}}

\theoremstyle{plain}
\newtheorem{Th}{Theorem}

\newtheorem{Cor}{Corollary}
\theoremstyle{definition}

\title[Resonances for ``large'' ergodic systems: a review]{Resonances
  for ``large'' ergodic systems in one dimension: a review}
\author{F. Klopp} \address[Fr{\'e}d{\'e}ric Klopp]{LAGA, U.M.R. 7539 C.N.R.S,
  Institut Galil{\'e}e, Universit{\'e} de Paris-Nord, 99 Avenue J.-B.
  Cl{\'e}ment, F-93430 Villetaneuse, France}
\email{\href{mailto:klopp@math.univ-paris13.fr}{klopp@math.univ-paris13.fr}}
\thanks{The author is supported by the grant ANR-08-BLAN-0261-01.}
\begin{document}
\begin{abstract}
  The present note reviews recent results on resonances for
  one-dimen\-sional quantum ergodic systems constrained to a large
  box. We restrict ourselves to one dimensional models in the discrete
  case. We consider two type of ergodic potentials on the half-axis,
  periodic potentials and random potentials. For both models, we
  describe the behavior of the resonances near the real axis for a
  large typical sample of the potential. In both cases, the linear
  density of their real parts is given by the density of states of the
  full ergodic system. While in the periodic case, the resonances
  distribute on a nice analytic curve (once their imaginary parts are
  suitably renormalized), in the random case, the resonances (again
  after suitable renormalization of both the real and imaginary parts)
  form a two dimensional Poisson cloud.
\end{abstract}
\setcounter{section}{-1}
\maketitle
\section{Introduction}
\label{sec:introduction}
\begin{figure}[h]
  \centering
  \includegraphics[height=1.5cm]{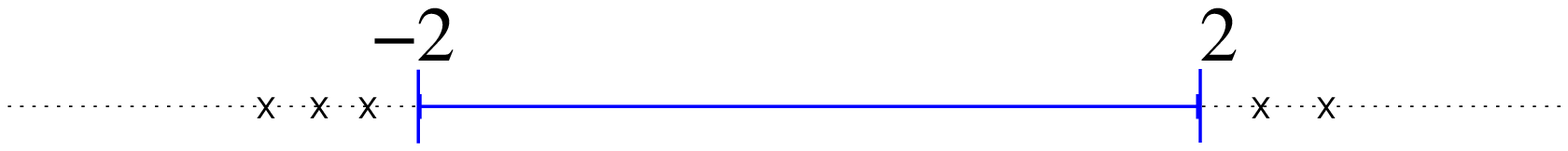}
  \hskip1cm
  \includegraphics[height=1.5cm]{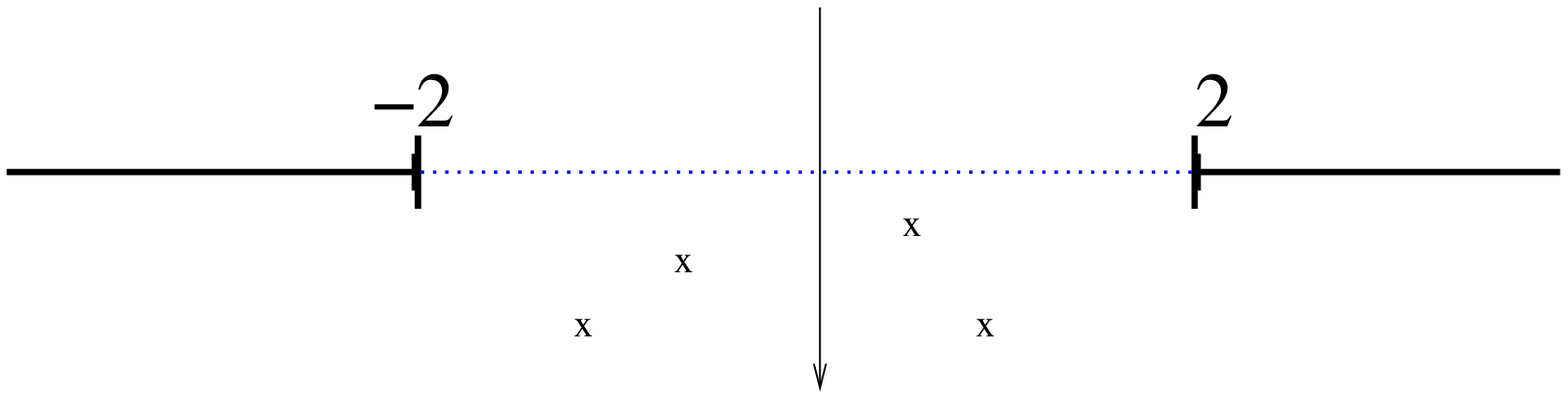}
  \caption{The spectrum of $H_L$ and the analytic continuation of
    $(z-H_L)^{-1}$}
  \label{fig:1}
\end{figure}

On $\ell^2(\N)$, consider $V$ a bounded potential and the operator
$H=-\Delta+V$ satisfying the Dirichlet boundary condition at $0$.\\
The potentials $V$ we will consider with are of two types:
\begin{itemize}
\item $V$ periodic;
\item $V=V_\omega$ random e.g. a collection of i.i.d. random
  variables.
\end{itemize}
The spectral theory of such models has been studied extensively (see
e.g.~\cite{MR2509110}) and it is well known that, when considered on
$\ell^2(\Z)$, the spectrum of $H$ is purely absolutely continuous when
$V$ is periodic (\cite{MR0650253}) while it is pure point when
$V=V_\omega$ is the Anderson potential
(\cite{MR1102675,MR94h:47068}). On $\ell^2(\N)$, the picture is the
same except for possible discrete eigenvalues outside the essential
spectrum which coincides and is of the same nature as the essential
spectrum of the operator on $\ell^2(\Z)$.\\
Let $L>0$. The object of our study is the following operator on
$\ell^2(\N)$
\begin{equation}
  \label{eq:2}
  H_L=-\Delta+V\car_{\llbracket 0,L\rrbracket}  
\end{equation}
when $L$ becomes large; here $-\Delta$ is the free Lapalce operator
defined by $-(\Delta u)(n)=u(n+1)+u(n-1)$ for $n\geq0$ where
$u=(u(n))_{n\geq0}\in\ell^2(\N)$ and $u(-1)=0$ (Dirichlet boundary
condition at $0$). \\
Clearly, the essential spectrum of $H_L$ is that of the discrete
Laplace operator, that is, $[-2,2]$, and it is absolutely
continuous. Moreover, outside this absolutely continuous spectrum,
$H_L$ has only discrete eigenvalues associated to exponentially
decaying eigenfunctions.\\
We are interested in the resonances of the operator $H_L$. These can
be defined as the poles of the meromorphic continuation of the
resolvent of $H_L$ through the continuous spectrum of $H_L$ (see
e.g.~\cite{MR1957536}). One proves that
\begin{Th}
  \label{thr:1}
  The operator valued holomorphic function $z\in\C^+\mapsto
  (z-H_L)^{-1}$ admits a meromorphic continuation from $\C^+$ to
  $\D\C\setminus\left((-\infty,2]\cup[2,+\infty)\right)$ (see
  Fig.~\ref{fig:1}) with values in the operators from
  $\ell^2_\text{comp}(\N)$ to $\ell^2_\text{loc}(\N)$.\\
  Moreover, the number of poles of this meromorphic continuation in
  the lower half-plane is equal to $L+1$.
\end{Th}
\noindent As said, we define the resonances as the poles of this
meromorphic continuation. The resonance widths, the imaginary part of
the resonances, play an important role in the large time behavior of
$\D e^{-itH_L}$, especially the smallest width that gives the leading
order contribution (see~\cite{MR1957536,MR1668841,MR1685889}).

\noindent As $L\to+\infty$, $H_L$ converges to $H$ in the strong
resolvent sense. Thus, it is natural to expect that the differences in
the spectral nature between the cases $V$ periodic and $V$ random
should reflect into differences in the behavior of the resonances. As
we shall see, this is the case. \\
Our goal is to describe the resonances or, rather, their statistical
properties and relate them (the distribution of the resonances, the
distribution of the widths) to the spectral characteristics of
$H=-\Delta+V$. In the periodic case, we expect that the Bloch-Floquet
data for the operator $-\Delta+V$ on $\ell(\Z)$ will be of importance;
in the random case, this role should be taken over by the distribution
of the eigenvalues of $-\Delta+V_\omega$.\vskip.2cm\noindent
The scattering theory or the closely related study of resonances for
the operator~\eqref{eq:2} or for similar one-dimensional models has
already been discussed in various works both in the mathematical and
physical
literature~\cite{MR1265236,MR1025232,MR2252707,1999PhRvL..82.4220T,MR1618595,PhysRevB.77.054203,MR1687454,MR2197681,PhysRevB.61.R2444,MR1042095}. The
proofs of the result we present below will be released elsewhere
(\cite{Kl:11c}). Though we will restrict ourselves to the discrete
model, the continuous model can be dealt with in a very similar
way.\vskip.2cm\noindent
Let us now describe our results. We start with the periodic case and
turn to the random case in the next section.
\section{The periodic case}
\label{sec:periodic-case}
\noindent We assume that, for some $p>0$, one has
\begin{equation}
  \label{eq:3}
  V_{n+p}=V_n\quad\text{for all}\quad n\geq0  .
\end{equation}
Let $\Sigma'$ be the spectrum of $H$ acting on $\ell^2(\N)$ and
$\Sigma_0$ be the spectrum of $-\Delta+V$ acting on $\ell^2(\Z)$. One
then has the following description for the spectra:
\begin{itemize}
\item $\displaystyle \Sigma'=\Sigma_0=\bigcup_{j=1}^p[a_j^-,a_j^+]$
  for some $a_j^-<a_j^+$ ($p\geq1$) and the spectrum is purely
  absolutely continuous (see e.g.~\cite{MR0650253}); the spectral
  resolution can be obtained via a Bloch-Floquet decomposition;
\item on $\ell^2(\N)$, one has (see e.g.~\cite{MR1301837})
  \begin{itemize}
  \item $\Sigma'=\Sigma_0\cup\{v_j; 1\leq j\leq n\}$ and $\Sigma_0$ is
    the a.c. spectrum of $H$;
  \item the $(v_j)_{0\leq j \leq n}$ are isolated simple eigenvalues
    associated to exponentially decaying eigenfunctions.
  \end{itemize}
\end{itemize}
When $L$ get large, it is natural to expect that the interesting
phenomena are going to happen near energies in $\Sigma'$. In
$\Sigma'\cap[(-\infty,-2)\cup(2,+\infty)$, one can check that $H_L$
has only discrete eigenvalues. We will now describe what happens for
the resonances near $[-2,2]$.
\subsection{The integrated density of states}
\label{sec:integr-dens-stat}
It is well known (see e.g.~\cite{MR94h:47068}) that one may define the
density of states of $H$, say $N(E)$, as the following limit
\begin{equation}
  \label{eq:4}
  N(E)=\lim_{L\to+\infty}\frac{\#\{\text{eigenvalues of
    }H_{\llbracket 0,L\rrbracket}\text{ in }(-\infty,E]\}}{L+1}.
\end{equation}
The restriction $H_{\llbracket 0,L\rrbracket}$ may be considered with
any boundary condition at $L$. The limit $N(E)$ defines a non
decreasing continuous function satisfying
\begin{itemize}
\item $N$ is real analytic and strictly increasing on
  $\overset{\circ}{\Sigma}_0$,
\item $N$ is constant outside $\Sigma_0$,
\item one has $N(-\infty)=0$ and $N(+\infty)=1$.
\end{itemize}
Thus, $dN$ defines a probability measure supported on $\Sigma_0$.
\subsection{Resonance free regions}
\label{sec:reson-free-regi}
We start with a description of the resonance free region.
\begin{Th}
  \label{thr:5}
  Let $I$ be a compact interval in $(-2,2)$.  Then,
  \begin{itemize}
  \item if $I\subset\R\setminus\Sigma'$, then, there exists $C>0$ such
    that, for $L$ sufficiently large, there are no resonances in
    $\{\text{Re}\,z\in I,\ \text {Im}\,z\geq-1/C\}$;
  \item if $I\subset\Sigma_0$, then, there exists $C>0$ such that, for
    $L$ sufficiently large, there are no resonances in
    $\{\text{Re}\,z\in I,\ \text {Im}\,z\geq-1/(CL)\}$;
  \item if $\{v_j\}=\overset{\circ}{I}\cap\Sigma'=I\cap\Sigma'$
    and $I\cap\Sigma_0=\emptyset$, then, for $L$ sufficiently large,
    there exists a unique resonance in $\{\text{Re}\,z\in I,\ \text
    {Im}\,z\geq-1/C\}$; moreover, this resonance, say $z_j$,
    satisfies, for some $\rho_j$ independent of $L$,
    \begin{equation}
      \label{eq:1}
      \text{Im} \,z_j\asymp - e^{-\rho_j L}\quad\text{ and }\quad
      |z_j-v_j|\asymp e^{-\rho_j L}.
    \end{equation}
  \end{itemize}
\end{Th}
\noindent So, below the spectral interval $(-2,2)$, except at the
discrete spectrum of $H$, there exists a resonance free region of
width at least of order $L^{-1}$. Each discrete eigenvalue of $H$
generates a resonance that is exponentially close to the real axis.
\subsection{Description of the resonances closest to $\Sigma_0$}
\label{sec:descr-reson}
\begin{wrapfigure}{r}{.42\textwidth}
  \begin{center}
    \includegraphics[width=.4\textwidth]{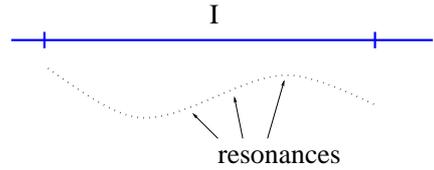}
  \end{center}
  \caption{The rescaled resonances for a periodic potential}
  \label{fig:2}
\end{wrapfigure}

Let $I$ be a compact interval in $(-2,2)\cap\overset{\circ}{\Sigma}_0$.\\
For $E\in\overset{\circ}{\Sigma}_0$, define
\begin{equation*}
  S(E)=\text{p.v.}
  \left(\int_\R\frac1{\lambda-E}dN(\lambda)\right)
  :=\lim_{\varepsilon\downarrow0}\int_{(-\infty,E_0-\varepsilon]
    \cup[E_0+\varepsilon,+\infty)}\frac1{\lambda-E}dN(\lambda).
\end{equation*}
The existence and regularity of the Cauchy principal value $S$ is
guaranteed by the regularity of $dN$ in $\overset{\circ}{\Sigma}_0$
(see e.g.~\cite{MR2542214}).\\
Let $(\lambda_j)_j=(\lambda_j^L)_j$ be the Dirichlet eigenvalues of
$(-\Delta+V)_{|\llbracket 0,L\rrbracket}$ in increasing order
(see~\cite{MR0650253}). We then prove the
\begin{Th}
  \label{thr:2}
  There exists $C_0>0$ such that, for $C>C_0$, there exists $L_0>0$
  such that for $L>L_0$, for $\lambda_j\in I$ such that
  $\lambda_{j+1}\in I$, there exists a unique resonance in
  $[\lambda_j,\lambda_{j+1}]+i[-CL^{-1},0]$, say $z_j$.  It satisfies
  \begin{equation*}
    z_j=\lambda_j+\frac{f(\lambda_j)}{L}\cot^{-1}\left(
      \left[e^{-i\arccos(\lambda_j/2)}+S(\lambda_j)\right]g(\lambda_j)\right)
    +o\left(\frac1L\right)
  \end{equation*}
  where $f$ and $g$ are real analytic functions defined by the Floquet
  theory of $H$ on $\Z$.
\end{Th}
\noindent The functions $f$ and $g$ can be computed explicitly in
terms of the Floquet reduction (see~\cite{Kl:11c}). Moreover, from
this quite explicit description of the resonances, one shows that, in
$I+i[-C/L,0]$, for $L$ sufficiently large,
\begin{itemize}
\item the resonances when rescaled to have imaginary parts of order
  $1$ accumulate on a real analytic curve;
\item the local (linear) density of resonances is given by the density
  of states of $H$.
\end{itemize}
More precisely, one proves
\begin{Cor}
  \label{cor:1}
  Fix $I$ as above. Then, there exists $C_0>0$, $V\supset I$ a
  neighborhood of $I$ and $h$ real analytic on $V$ such that, for
  $C>C_0$, there exists $L_0>0$ such that for $L>L_0$,
  \begin{itemize}
  \item if $z\in I+i[-CL^{-1},0]$ is a resonance of $H_L$, then
    \begin{equation}
      \label{eq:6}
      L\cdot\text{Im}\,z=h(\text{Re}\,z)+o(1);
    \end{equation}
  \item for $J\subset I$, any interval one has
    \begin{equation}
      \label{eq:7}
      \frac{\#\{z\in J+i[-CL^{-1},0],z\text{ resonance of
        }H_L\}}{L+1}=\int_J dN(E)+o(1).
    \end{equation}
  \end{itemize}
\end{Cor}
\noindent Fig.~\ref{fig:2} pictures the resonances after rescaling
their width by $L$: these are nicely spaced points interpolating a
smooth curve.
\subsection{Description of the low lying resonances}
\label{sec:descr-reson-ll}
One can also study what happens below the lines Im$z=-C/L$. Therefore,
one considers the function
\begin{equation}
  \label{eq:8}
  \hskip-5cm\Xi(z):=\int_\R\frac{dN(\lambda)}{\lambda-z}+e^{-i\arccos(z/2)}
\end{equation}
defined in the lower half-plane where, the function $z\mapsto\arccos
z$ is the analytic continuation to the lower half-plane of the
determination taking values in $[-\pi,0]$ over the interval
$[-1,1]$.\\
A study of the behavior of this function on the boundary
$\partial\C^-$ shows that, on $\C^-$, $\Xi$ admits at most a finite
number of zeros that are all simple. One proves
\begin{Th}
  \label{thr:3}
  There exists $C_0>0$ and $L_0>0$ such that, for $L>L_0$, 
  \begin{itemize}
  \item the number of resonances of $H_L$ in the half-plane
    $\R+i(-\infty,-C_0/L]$ is equal to the number of zeros of $\Xi$;
  \item to each zero of $\Xi$, say $\tilde z$, one can associate a
    unique resonance, say $z$, that satisfies $|z-\tilde z|\lesssim
    L^{-1}$.
  \end{itemize}
\end{Th} 
\noindent So, by Theorem~\ref{thr:2}, except for a finite number of
resonances that converge to the zeros of $\Xi$, all the resonances of
$H_L$ converge to the spectrum of $H$.
\section{The random case}
\label{sec:random-case}
Let now $V=V_\omega$ where $V_\omega(n)=\omega_n$ and
$(\omega_n)_{n\geq0}$ are bounded independent and identically
distributed random variables. Assume that the common
law of the random variables admits a bounded density, say, $g$.\\
Set $H_\omega=-\Delta+V_\omega$ on $\ell^2(\N)$. Let
$\sigma(H_\omega)$ be the spectrum of $H_\omega$ and $\Sigma$ be the
almost sure spectrum of $-\Delta+V_\omega$ acting on $\ell^2(\Z)$
(see~\cite{MR2509110}); one knows that
\begin{equation*}
  \Sigma=[-2d,2d]+\text{supp}\,g
\end{equation*}
One has the following description for the spectra:
\begin{itemize}
\item $\omega$-almost surely, $\sigma(H_\omega)=\Sigma$; the spectrum
  is purely punctual; it consists of simple eigenvalues associated to
  exponentially decaying eigenfunctions (Anderson localization, see
  e.g.~\cite{MR94h:47068,MR2509110}); one can prove that the whole
  spectrum is dynamically localized;
\item on $\ell^2(\N)$, one has (see e.g.~\cite{MR94h:47068,MR1102675})
  \begin{itemize}
  \item $\omega$-almost surely, $\sigma(H_\omega)=\Sigma\cup
    K_\omega$;
  \item $\Sigma$ is the essential spectrum of $H_\omega$; it consists
    of simple eigenvalues associated to exponentially decaying
    eigenfunctions;
  \item the set $K_\omega$ is the discrete spectrum of $H_\omega$
    which may be empty and depends on $\omega$.
  \end{itemize}
\end{itemize}
\subsection{The integrated density of states and the Lyapunov
  exponent}
\label{sec:lyapunov-exponent}
The integrated density of states is defined by~\eqref{eq:4}. It is the
distribution function of a probability measure supported on
$\Sigma$. As the common law of the random variables
$(\omega_n)_{n\geq0}$ admits a bounded density, the integrated density
of states $N(E)$ is known to be Lipschitz continuous
(\cite{MR94h:47068,MR2509110}). Let $\displaystyle
n(E)=\frac{dN}{dE}(E)$ be its derivative; it exists for almost every
$E$.\\
One also defines the Lyapunov exponent, say $\rho(E)$ as follows
\begin{equation}
  \label{eq:9}
  \rho(E)=\lim_{L\to+\infty}\frac1{L+1}\log
  \left\|\prod_{n=L}^0 \begin{pmatrix} E-V_\omega(n) & -1
      \\ 1 &0    \end{pmatrix}  \right\|.
\end{equation}
For any $E$, $\omega$-almost surely, this limit is known to exist and
to be independent of $\omega$ (see
e.g.~\cite{MR94h:47068,MR1102675}). Moreover, it is positive and
continuous for all $E$ and the Thouless formula states that it is the
harmonic conjugate of $n(E)$ (see e.g.~\cite{MR883643}).
\subsection{Resonance free regions}
\label{sec:reson-free-regi1}
We again start with a description of the re\-sonance free region near
the spectrum of $-\Delta$. As in the periodic case, the size of this
region will depend on whether an energy belongs to the essential
spectrum of $H_\omega$ or not. We prove
\begin{Th}
  \label{thr:4}
  Let $I$ be a compact interval in $(-2,2)$. Then, one has
  \begin{itemize}
  \item there exists $C>0$ such that, $\omega$-a.s., if
    $I\subset\R\setminus\sigma(H_\omega)$, then, for $L$ sufficiently
    large, there are no resonances of $H_{\omega,L}$ in
    $\{\text{Re}\,z\in I,\ \text {Im}\,z\geq-1/C\}$;
  \item there exists $C>0$ such that, $\omega$-a.s., if
    $\{v_j\}=\{v_j(\omega)\}=\overset{\circ}{I}\cap K_\omega=I\cap
    K_\omega$ and $I\cap\Sigma=\emptyset$, then, for $L$ sufficiently
    large, there exists a unique resonance in $\{\text{Re}\,z\in I,\
    \text {Im}\,z\geq-1/C\}$; moreover, this resonance, say $z_j$,
    satisfies~\eqref{eq:1} for some $\rho_j=\rho_j(\omega)$
    independent of $L$.
  \item if $I\subset\overset{\circ}{\Sigma}$, then, there exists $C>0$
    such that, $\omega$-a.s., for $L$ sufficiently large, there are no
    resonances of $H_{\omega,L}$ in $\{\text{Re}\,z\in I,\ \text
    {Im}\,z\geq -e^{-2\rho L(1+o(1))})\}$ where $\rho$ is the maximum
    of the Lyapunov exponent $\rho(E)$ on $I$.
  \end{itemize}
\end{Th}
\noindent When comparing this result with Theorem~\ref{thr:5}, it is
striking that the width of the resonance free region below $\Sigma$ is
much smaller in the random case than in the periodic case. This a
consequence of the localized nature of the spectrum i.e. of the
exponential decay of the eigenfunction.
\subsection{Description of the resonances close to $\Sigma$}
\label{sec:descr-reson1}
\begin{wrapfigure}{r}{.42\textwidth}
  \begin{center}
    \includegraphics[height=2cm]{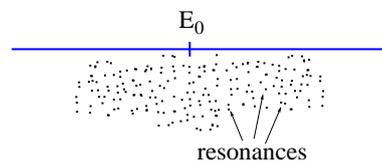}
  \end{center}
  \caption{The rescaled resonances for a random potential}
  \label{fig:3}
\end{wrapfigure}

We will now see that below the resonance free strip exhibited in
Theorem~\ref{thr:4} one does find resonances, actually, many of
them. We prove
\begin{Th}
  \label{thr:6}
  Let $I$ be a compact interval in $(-2,2)\cap\overset
  {\circ}{\Sigma}$.  Then, $\omega$-a.s.,
  \begin{itemize}
  \item for any $\kappa\in(0,1)$, one has
    \begin{equation*}
      \frac1L\#\left\{z\text{ resonance of }H_{\omega,L}\text{ s.t. Re}\,z\in I,\
        \text{Im}\,z\geq -e^{-L^\kappa}\right\}\to \int_IdN(E);
    \end{equation*}
  \item fix $E\in I$ such that $n(E)>0$; then, for $\delta>0$, there
    exits $\varepsilon>0$ such that
    \begin{equation*}
      \liminf_{L\to+\infty}\frac1L\#\left\{\text{resonances }z
        \text{ s.t. }
        \begin{split}
        \text{ Re}\,z\in[E-\varepsilon,E+\varepsilon],\\
        \text{Im}\,z\geq -e^{-2(\rho(E)-\delta)L}   
        \end{split}
      \right\}>0.
    \end{equation*}
  \end{itemize}
\end{Th}
\noindent The first striking fact is that the resonances are much
closer to the real axis than in the periodic case; the lifetime of
these resonances is much larger. The resonant states are quite stable
with lifetimes that are exponentially large in the width of the random
perturbation.\\
The structure of the set of resonances is also very different from the
one observed in the periodic case (see Fig.~\ref{fig:2}) as we will
see now. Let $I$ be a compact interval in $(-2,2)\cap\overset
{\circ}{\Sigma}$ and $\kappa\in(0,1)$. Fix $E_0\in I$ such that
$n(E_0)>0$.\\
Let $(z_i^L(\omega))_i$ be the resonances of $H_{\omega,L}$ in $\D
K_L:=[E_0-\varepsilon,E_0+\varepsilon]+i\left[-e^{-L^\kappa},0\right]$. We
first rescale the resonances: define
\begin{equation}
  \label{eq:10}
  \begin{aligned}
    x_j&=x_j^L(\omega)=n(E_0)\,L\,(\text{Re}\,z_j^L(\omega)-E_0)\\
    y_j&=y_j^L(\omega)=-\frac{1}{2\rho(E_0)\,L}\log|\text{Im}\,z_j^L(\omega)|.
  \end{aligned}
\end{equation}
Let us note that the scaling of the real and of the imaginary of the
resonances are very different. According to the conclusions of
Theorem~\ref{thr:6}, this scaling essentially sets the mean spacing
between the real parts of the resonances to $1$ and the imaginary
parts to be of order $1$.\vskip.1cm\noindent
Consider now the two-dimensional point process $\xi_L(E_0,\omega)$
defined
\begin{equation}
  \label{eq:11}
  \xi_L(E_0,\omega)=\sum_{z_j^L\in K_L}\delta_{(x_j,y_j)}.  
\end{equation}
We prove
\begin{Th}
  \label{thr:7}
  The point process $\xi_L$ converges weakly to a Poisson process in
  $\R\times[0,1]$ with intensity $1$. That is, for any $p\geq0$, if
  $(I_n)_{1\leq n\leq p}$ resp. $(C_n)_{1\leq n\leq p}$, are disjoint
  intervals of the real line $\R$ resp. of $[0,1]$, then
  \begin{equation*}
    \lim_{L\to+\infty}
    \pro\left(\left\{\omega;\
        \begin{aligned}
          &\#\left\{j;
            \begin{aligned}
              x_j(\omega,\Lambda)&\in I_1 \\ y_j(\omega,\Lambda)&\in
              C_1 \end{aligned}
          \right\}=k_1\\&\hskip1cm\vdots\hskip2cm\vdots\\
          &\#\left\{j;
            \begin{aligned}
              x_j(\omega,\Lambda)&\in I_p \\ y_j(\omega,\Lambda)&\in C_p
            \end{aligned}
          \right\}=k_p
        \end{aligned}
      \right\}\right)=
    \prod_{n=1}^pe^{-\mu_n}
    \frac{(\mu_n)^{k_n}}{k_n!},
  \end{equation*}
  where $\mu_n:=|I_n||C_n|$ for $1\leq n\leq p$.
\end{Th}
\noindent Hence, after rescaling the picture of the resonances (see
Fig~\ref{fig:3}) is that of points chosen randomly independently of
each other in $\R\times[0,1]$.
This is the analogue of the celebrated result on the Poisson
structure of the eigenvalues for a random system
(see e.g.~\cite{MR84e:34081,MR97d:82046,Ge-Kl:10})
\vskip.2cm\noindent In~\cite{Kl:10}, we proved decorrelation estimates
that can be used in the present setting to prove
\begin{Th}
  \label{thr:8}
  Fix $E_0\not=E'_0$ such that $n(E_0)>0$ and $n(E'_0)>0$. Then, the
  limits of the processes $\xi_L(E_0,\omega)$ and $\xi_L(E'_0,\omega)$
  are stochastically independent.
\end{Th}
\noindent Due to the rescaling, the above results give a picture of
the resonances in a zone of the type
\begin{equation*}
  E_0+L^{-1}\left[-\varepsilon^{-1},\varepsilon^{-1}\right]
  -i\,\left[e^{-2(1-\varepsilon)\rho(E_0)L}
    ,e^{-2\varepsilon\rho(E_0)L}\right] 
\end{equation*}
For $\varepsilon>0$ small fixed, when $L$ gets large, this rectangle
is of a very small width and located very close to the real axis. One
can actually bet a picture of the resonances near a line
Im$z=e^{-2\rho(E_0)L\delta}$ for $\delta\in[0,1]$. Though the
resonances closest to the real axis play the most important role in
scattering (see e.g.~\cite{MR1957536}), such a picture answers the
natural question of what happens deeper in the lower half-plane.\\
Fix an increasing sequence of scales $\ell=(\ell_L)_L$ such that
\begin{equation}
  \label{eq:14}
  \frac{\ell_L}{\log L}\vers_{L\to+\infty}+\infty
  \quad\text{ and }\quad \frac{\ell_L}L\vers_{L\to+\infty}0.
\end{equation}
Fix $x_0\in[0,1]$ and $E_0\in I$ so that $\nu(E_0)>0$. Let
$(z_i^L(\omega))_i$ be the resonances of $H_{\omega,L}$ in $\D \tilde
K_L:=E_0+\ell_L^{-1}\left[-\varepsilon^{-1},\varepsilon^{-1}\right]+
i\left[-e^{-\ell_L},0\right]$. Note that $\tilde K_L$ is much larger
than $K_L$.  We first rescale the resonances using the sequence
$(\ell_L)_L$: define
\begin{equation}
  \label{eq:12}
  \begin{aligned}
    x_j&=x_j^{\ell_L}(\omega)=n(E_0)\,\ell_L\,(\text{Re}\,z_j^L(\omega)-E_0),\\
    y_j&=y_j^{\ell_L}(\omega)=-\frac1{2\,\rho(E_0)\,\ell_L}\left[
      2\,\rho(E_0)\,L\,x_0 +\log|\text{Im}\,z_j^L(\omega)|\right].
  \end{aligned}
\end{equation}
Note that if $\ell_L=L$, we recover~\eqref{eq:10} up to a shift for
the imaginary part. The scaling introduced in~\eqref{eq:12} is
different from the one introduced in~\eqref{eq:10}. Here, we look at a
window of size $\ell^{-1}_L$ around $E_0$ is the real parts (it is
thus much larger than the $L^{-1}$ window considered
in~\eqref{eq:10}); but, in the imaginary part, on a logarithmic scale
we look at a window of size $\ell_L$ around the point $-2\rho(E_0)x_0
L$ (this is much smaller than the window of size $L$ considered
in~\eqref{eq:10}-\eqref{eq:11}). This covariant scaling of the real
and imaginary parts of the resonances is the correct analogue of the
covariant scaling introduced in~\cite{Ge-Kl:10} for the eigenvalues
and localization centers of a random operator in the localized phase.
\vskip.1cm\noindent
Consider now the two-dimensional point process
$\xi_{L,\ell}(x_0,E_0,\omega)$ defined by
\begin{equation}
  \label{eq:13}
  \xi_{L,\ell}(x_0,E_0,\omega)=\sum_{z_j^L\in K_L}\delta_{(x_j,y_j)}.  
\end{equation}
We prove
\begin{Th}
  \label{thr:9}
  For $x_0\in[0,1]$ and $E_0\in I$ so that $\nu(E_0)>0$, the point
  process $\xi_L(x_0,E_0,\omega)$ converges weakly to a Poisson
  process in $\R\times\R$ with intensity $1$.
\end{Th}
\noindent So we see that the local picture of the rescaled resonances
around the point $E_0$ near the line Im$z=-ie^{-2\rho(E_0)x_0L}$ is
essentially independent of $E_0$ and $x_0$. A value of particular
interest is $x_0=0$. In this case, we look at resonances lying at a
distance of order $e^{-\ell_L}$ from the real axis (thus, much further
away that the resonances considered in Theorem~\ref{thr:7}). By
condition~\eqref{eq:14}, we can get a precise description of the
resonances lying almost at a polynomial distance to the real
axis.\vskip.1cm\noindent
For the processes $(\xi_L(x_0,E_0,\omega))_{x_0,E_0}$, one gets an
asymptotic independence result analogous to Theorem~\ref{thr:8},
namely,
\begin{Th}
  \label{thr:10}
  Fix $E_0$ and $E'_0$ such that $n(E_0)>0$ and $n(E'_0)>0$ and $x_0$
  and $x'_0$ such that $(x_0,E_0)\not=(x'_0,E'_0)$.\\ 
  Then, the limits of the processes $\xi_L(x_0,E_0,\omega)$ and
  $\xi_L(x'_0,E'_0,\omega)$ are stochastically independent.
\end{Th}
\noindent One can get a number of other statistics using the
techniques developed for the study of the spectral statistics of the
eigenvalues of a random system in the localized phase
(see~\cite{Ge-Kl:10b,Ge-Kl:10}).
\subsection{The description of the low lying resonances}
\label{sec:descr-low-lying}
One can also study what happens below the lines Im$z=-e^{-\log^\alpha
  L}$ (for $\alpha>1$). As we will see the picture is quite different
from that obtained for the periodic case (see
section~\ref{sec:descr-reson-ll}). \\
To state the result, it will be convenient to change the notation
slightly and to write the operator $H_{\omega,L}$ as
\begin{equation}
  \label{eq:16}
  \tilde H_{\omega,L}=-\Delta+\tilde V_{\omega,L}\text{ where }V_{\omega,L}=
  \begin{cases}
    \omega_{L-n}&\text{ if }n\in\llbracket 0,L\rrbracket,\\ 0&\text{
      if not}.
  \end{cases}
\end{equation}
As the random variables $(\omega_n)_{n\geq0}$ are i.i.d, the families
$(H_{\omega,L})_\omega$ and $(\tilde H_{\omega, L})_\omega$ have the
same distribution.\\
On $\ell^2(\Z)$, consider the following random operator
\begin{equation}
  \label{eq:17}
  \tilde H_\omega=-\Delta+\tilde V_\omega\text{ where }V_{\omega}=
  \begin{cases}
    \omega_{-n}&\text{ if }n\leq 0,\\ 0&\text{ if }n>0.
  \end{cases}
\end{equation}
The resonances (in the sense of Theorem~\ref{thr:1}) of this operator
were studied in~\cite{PhysRevB.77.054203,MR2252707}. It was in
particular shown that, below any line Im$\, z=-1/C$, $\tilde H_\omega$
has at most finitely many resonances.\\
We show
\begin{Th}
  \label{thr:11}
  Fix $\alpha>1$. Then, $\omega$-almost surely, there exists $L_0>0$
  and $C_0>0$ such that, for $L>L_0$,
  \begin{itemize}
  \item the number of resonances of $\tilde H_{\omega, L}$ in the
    half-plane $\D\R+i\left(-\infty, -e^{-\log^\alpha L}\right]$ is
    equal to the number of resonances of $\tilde H_\omega$ in the same
    half-plane;
  \item to each resonance of $\tilde H_\omega$ in this half-plane, say
    $\tilde z$, one can associate a unique resonance of $\tilde
    H_{\omega,L}$, say $z$, that satisfies $|z-\tilde z|\leq
    e^{-L/C_0}$.
  \end{itemize}
\end{Th}
\noindent Hence, we see that, in opposition to the periodic case, not
all resonances except finitely many of them converge to the
spectrum $H_\omega$.\\
It is interesting to note that the result of Theorem~\ref{thr:11}
stays true on the (larger) half-plane $\D\R+i\left(-\infty,
  -e^{-L^\beta}\right]$ (for $\beta\in(0,1)$).\\
Finally, let us say that one can combine Theorem~\ref{thr:9}
and~\ref{thr:11} to recover the results
of~\cite{PhysRevB.77.054203,MR2252707} on behavior of the average
density of resonances of $\tilde H_\omega$ at a distance $y$ from the
real axis in the limit $y\to0$.
\def\cprime{$'$} \def\cydot{\leavevmode\raise.4ex\hbox{.}} \def\cprime{$'$}

\end{document}